\documentclass[prl,amsmath,amssymb,superscriptaddress,twocolumn]{revtex4}
\usepackage{graphicx}

\newcommand{\ba}{{\bf a}}
\newcommand{\br}{{\bf r}}

\newcommand{\bk}{{\bf k}}

\newcommand{\eps}{\epsilon}

\DeclareMathAlphabet{\mathpzc}{OT1}{pzc}{m}{it} 
\begin{document}
\title{Dynamical conductivity of ungated suspended graphene}
\author{Oskar Vafek}
\affiliation{National High Magnetic Field Laboratory and Department
of Physics, Florida State University, Tallahassee, Florida 32306,
USA}
\address{}
\date{\today}
\begin{abstract}
Frequency dependent conductivity of Coulomb interacting massless
Dirac fermions coupled to random scalar and random vector potentials
is found as a function of frequency in the regime controlled by a
line of fixed points. Such model provides a low energy description
of a weakly rippled suspended graphene. The main finding is that at
the neutrality point the a.c. conductivity is not frequency
independent and may either increase or decrease with decreasing
$\omega$, depending on the values of the disorder variances
$\Delta_{\phi}$, $\Delta_{A}$ and the Coulomb coupling
$\alpha=e^2/(\eps v_F)$. The low frequency behavior is characterized
by the values of two dimensionless parameters
$\gamma=\Delta_{\phi}/\alpha^2$ and $\Delta_A$ which are RG
invariants, and for small values of which the electron-hole
"puddles" are effectively screened making the results asymptotically
exact.
\end{abstract}

 \maketitle
The physics of massless Dirac fermions in two spatial dimensions has
received renewed attention since their discovery in single-layer
graphene\cite{Novoselov:2005fk,Zhang:2005uq,Geim:2007lr}. The great
interest is not unrelated to the quantum
critical\cite{Gonzales1994,herbutPRL06,vafek:prl2007,sheehy:226803,HerbutJuricicVafekPRL2008,fritz:085416}
nature of the system near the neutrality point, where the Fermi
level lies precisely at the (Dirac-like) band crossing. Indeed,
absence of any intrinsic long distance lengthscale sets constraints
on the (low) frequency or temperature dependence of any physical
quantity. In this regard, electrical conductivity $\sigma$ plays a
special role since in two spatial dimensions it is expected to be
proportional to $e^2/h$; the proportionality constant, which need
not be finite, depends only on the nature of the renormalization
group (RG) fixed point characterizing the low energy-long distance
physics\cite{Fisher1990,Cha1991,Damle97,HerbutJuricicVafekPRL2008,sheehy:226803}.
Electrical conductivity measurements at the neutrality point
therefore constitute a direct probe of the non-trivial physics
emerging at the end of the RG trajectory.

 Recent experiments performed on monolayer graphene, both
suspended\cite{Nair06062008} and on the substrate
\cite{basov2008,dawlaty-2008}, have found that the optical
conductivity near the neutrality point is
$\sigma(\omega)=\frac{\pi}{2}\frac{e^2}{h}$, i.e. largely frequency
independent and equal to $\pi/2$ it the natural units. What little
frequency dependence there is in the regime where $\hbar \omega\sim
1eV$ can be attributed to the curvature corrections to the
electronic dispersion which deviates from the perfectly conical
massless Dirac-like at such large energies\cite{Nair06062008}. While
presently there is no conductivity data at the lower frequencies of
interest (at sub $meV$ scales), it is natural to ask whether such
frequency independent $\sigma(\omega)$ should persist down to
$\hbar\omega\sim k_BT$. Since the role of charged impurities located
at the substrate is naturally eliminated in the suspended samples,
the dominant source of scattering is most likely the random
configuration of strain due to the graphene sheet rippling and
possibly from the boundary effects imposed by the scaffolding
necessary for the actual suspension. Such long wavelength strain
fields are known to couple to the massless Dirac particles of
graphene as a vector potential and a scalar
potential\cite{suzuura,manes:045430,mariani:076801}, and their
combined effect on the a.c. conductivity, together with the effects
of the electron-electron (Coulomb) interactions, are analyzed below.

In the non-interacting model of massless Dirac particles coupled to
the random scalar and random vector potentials with variances
$\Delta_{\phi}$ and $\Delta_A$, it has been long known that within
the perturbative RG, $\Delta_{\phi}$ grows upon approaching low
energies and the theory flows to a perturbatively inaccessible fixed
point \cite{Ludwig:prb94,aleinerEfetov2006}. The effects of Coulomb
interactions, parameterized by a dimensionless coupling
$\alpha=e^2/\eps v_F$, and random vector potential (without scalar
potential randomness) on the a.c. conductivity has been studied in
Ref.\cite{HerbutJuricicVafekPRL2008} where it was found that the
a.c. conductivity is non-universal and dependent only on $\Delta_A$
which is marginal in the RG sense.  Such non-universality is
directly tied to the appearance of the infra-red (IR) locally stable
line of fixed points. The two loop effects on the RG flow diagram
have been incorporated in Ref.\cite{VafekCasePRB2008} and extended
to other types of disorder in Ref\cite{fosterAleinerPRB2008}.

\begin{figure}[t]
\begin{center}
\begin{tabular}{c}
\includegraphics[width=0.35\textwidth]{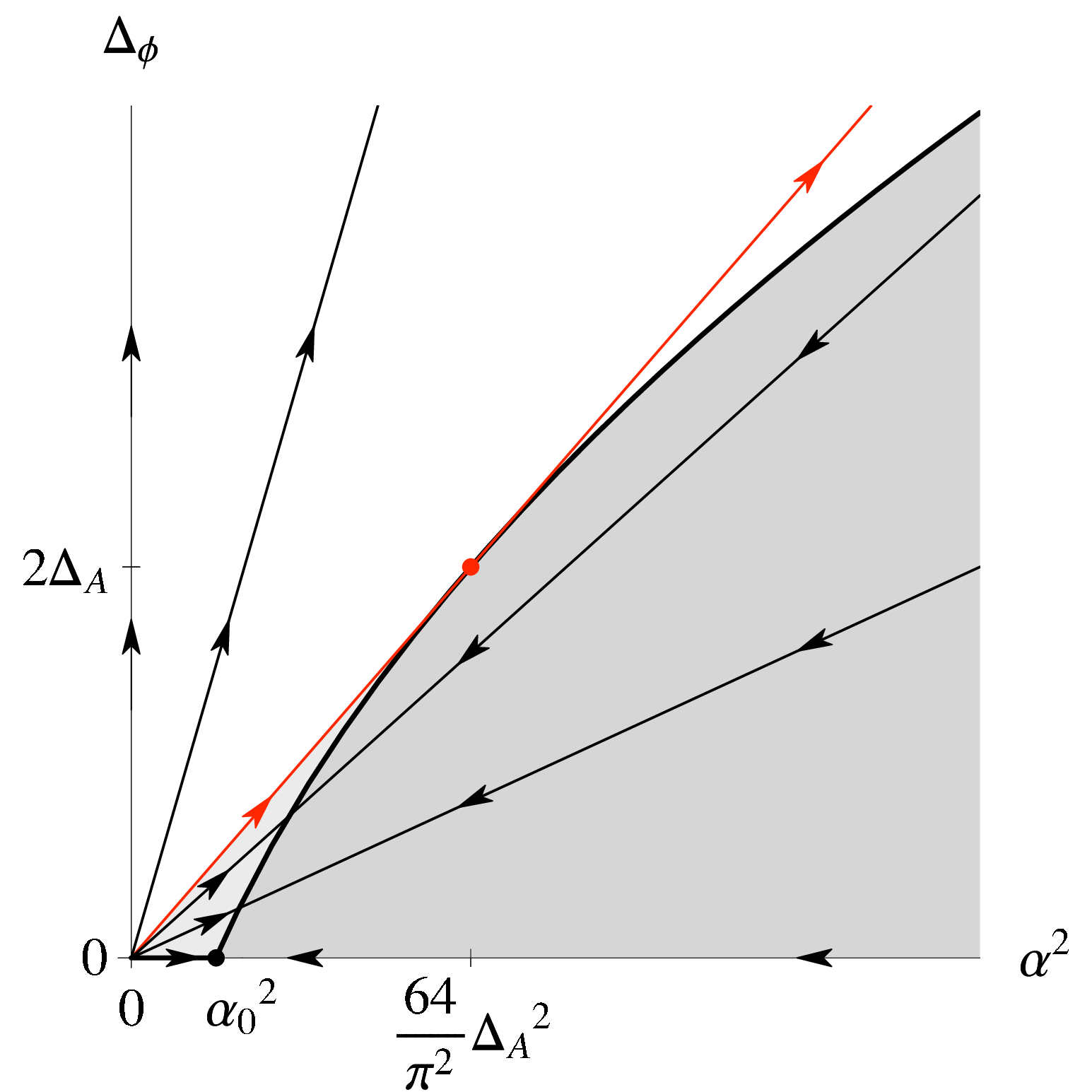}
\end{tabular}
\end{center}
\caption{The renormalization group flow diagram in the scalar
disorder $\Delta_{\phi}$ -- Coulomb interaction $\alpha=e^2/(\eps
v_F)$ plane. There are two marginal parameters (RG invariants): the
variance of the random vector potential $\Delta_A$ and the ratio
$\gamma=\Delta_{\phi}/\alpha^2$. For $\gamma<32/(\pi^2\Delta_A)$
each RG trajectory can cross three fixed points (the strong coupling
one is not shown). The middle one, which is IR stable, merges with
the IR unstable at a multicritical point (red circle) when
$\gamma=32/(\pi^2\Delta_A)$. $\alpha_0=4\Delta_A/\pi$ lies along the
fixed line discussed in
\cite{HerbutJuricicVafekPRL2008,VafekCasePRB2008}. For
$\gamma>32/(\pi^2\Delta_A)$, there are runaway flows with no
perturbatively accessible fixed points. The phase diagram splits
naturally into two regimes: Regime I (shaded) given by the locus of
points which eventually run into the perturbatively accessible IR
stable fixed line show above, and Regime II (unshaded) with runaway
flows. The $\omega\rightarrow 0$ limit of the (collisionless) a.c.
conductivity along the IR stable fixed line is given in
Eq.(\ref{eq:Conductivity}), and its $\omega$ dependence is discussed
in the text.} \label{fig:RGflows}
\end{figure}

In this work, we study the combined effects of the unscreened
Coulomb interactions, the quenched random vector and scalar
potential disorder which arise naturally in the model of (randomly)
strained sample, and analyze the frequency dependent conductivity
within perturbative RG. This includes the effects of monolayer
ripples and electron-hole "puddles". The perturbative RG adopted
here is technically much simpler than the large $N$ approximation
adopted by Foster and Aleiner\cite{fosterAleinerPRB2008} to map out
the phase diagram, and to leading order, leads to qualitatively
similar results (for differences beyond the leading order, and the
advantages of the former, see Ref.\cite{VafekCasePRB2008}).
Moreover, the weak coupling RG can be easily extended to the
calculation of $\sigma(\omega)$, which was not calculated in
\cite{fosterAleinerPRB2008}. The main finding is that
$\sigma(\omega)$ is not frequency independent and may either
increase or decrease with decreasing $\omega$, depending on the
values of $\Delta_{\phi}$, $\Delta_{A}$ and $\alpha$. The low
frequency behavior is characterized by the values of the
dimensionless parameters $\gamma=\Delta_{\phi}/\alpha^2$ and
$\Delta_A$ which are RG invariants. The ultimate low frequency
behavior depends on being in one of two regimes (see
Fig.\ref{fig:RGflows}).

Specifically, for the bare couplings in the Regime I, which is
determined by the conditions $\alpha<8\Delta_A/\pi$ {\it and}
$\gamma<\pi^2/(32\Delta_A)$ or $\alpha>8\Delta_A/\pi$ {\it and}
$\Delta_{\phi}<(\pi/2)\alpha-2\Delta$, then as $\omega\rightarrow 0$
\begin{eqnarray}
\sigma\label{eq:Conductivity}
\rightarrow\frac{e^2}{h}\left[\frac{\pi}{2}+\frac{\Delta_A}{6}+\frac{(23-6\pi)\pi^2}{96\gamma}\left(1-\sqrt{1-\frac{32}{\pi^2}\gamma\Delta_A}\right)\right].
\end{eqnarray}
subject to the constraint $0<\gamma<\pi^2/(32\Delta_A)$. In this
parameter regime the weak coupling RG flow equations lead to a {\it
perturbatively accessible} IR stable line of fixed points, making
the above result asymptotically exact. The ac conductivity, in the
collisionless limit of interest here, therefore behaves as a
universal amplitude, depending only on the RG invariants $\Delta_A$
and $\gamma$. In the Regime II, which encompasses the parameter
regime not included in Regime I there are no perturbatively
accessible fixed points, and the problem remains open. The details
of the calculation leading to the above claims, as well as $\omega$
dependence of the conductivity, are presented below.

We start with the imaginary time partition function
\begin{eqnarray}  Z=\int\mathcal{D}\bar{\psi}\psi
e^{-(S_0+S_{dis}+S_{int})}
\end{eqnarray}
where
\begin{eqnarray}  S_0&=&\!\!\!\int_0^{\beta}d\tau\int
d^2r \bar{\psi}(r,\tau)\left(\partial_{\tau}+v_F\sigma\cdot{\bf
p}\right)\psi(r,\tau)\\
S_{dis}&=&\!\!\!\int_0^{\beta}d\tau\int d^2r
\bar{\psi}(r,\tau)\left(\phi({\bf
r})+ v_F\sigma\cdot{\bf a}\right)\psi(r,\tau)\\
S_{int}&=&\!\!\!\frac{1}{2} \int_0^{\beta}\!\! d\tau \!\!\int\!\!
d^2r d^2r' \!\bar{\psi}\psi(r,\tau)V(|r-r'|)\bar{\psi}\psi(r',\tau)
\end{eqnarray}
The last term corresponds to the (Coulomb) electron-electron
interaction $V(|r-r'|)=\frac{e^2}{\eps|r-r'|}$, where $\eps$ is the
dielectric constant which may differ from $1$. We assume that the
disorder is uncorrelated with variances:
\begin{eqnarray}  \langle \phi_{{\bf
k}}\phi_{{\bf k'}}\rangle &=& (2\pi)^2\delta({\bf k}-{\bf
k'})v^2_F\Delta_{\phi}\\
\langle a^{\mu}_{{\bf k}}a^{\nu}_{{\bf k'}}\rangle &=&
(2\pi)^2\delta({\bf k}-{\bf k'})\delta_{\mu\nu}\Delta_{A}.
\end{eqnarray}
As has been discussed extensively in the past, the scalar and vector
potentials are naturally connected to the appearance of strain
tensor $u_{ij}$ as $\phi=g(u_{xx}+u_{yy})$, $a_x=b(u_{yy}-u_{xx})$,
$a_y=2bu_{xy}$\cite{suzuura,manes:045430,mariani:076801,guinea:205421},
with the estimates $g\approx 20-30eV$\cite{suzuura,mariani:076801}
and $b\approx \AA^{-1}$\cite{suzuura,guinea:205421}.

We can perform the (quenched) average over the gaussian disorder
fields $\phi(\br)$ and $\ba(\br)$ using the standard replica trick
of including $n$ copies of the fermion fields: $\psi\rightarrow
\psi^i$, where $i=1,2,\ldots,n$. The resulting replica field theory
is
\begin{eqnarray}  \langle
Z^n\rangle_{dis}=\int\mathcal{D}\bar{\psi}^i\psi^i
e^{-(S_0+S_{\phi}+S_{A}+S_{int})}
\end{eqnarray}
where
\begin{eqnarray}  S_0&=&\!\!\int_0^{\beta}d\tau\int
d^2r \bar{\psi}^i\left(\partial_{\tau}+v_F\sigma\cdot{\bf
p}\right)\psi^i \nonumber\\
S_{\phi}&=&\!\!-\frac{v^2_F\Delta_{\phi}}{2}\int_0^{\beta} d\tau
d\tau'\int
d^2{\bf r}\bar{\psi}^i\psi^i(r,\tau)\bar{\psi}^j\psi^j(r,\tau')\nonumber\\
S_{A}&=&\!\!-\frac{v^2_F\Delta_{A}}{2}\!\int_0^{\beta}\! d\tau
d\tau'\!\!\!\int d^2{\bf
r}\bar{\psi}^i\sigma^{\mu}\psi^i(r,\tau)\bar{\psi}^j\sigma^{\mu}\psi^j(r,\tau')\nonumber\\
S_{int}&=&\!\!\frac{1}{2}\! \int_0^{\beta}\!\! d\tau\!\! \int d^2r
d^2r'
\bar{\psi}^i\psi^i(r,\tau)V(|r-r'|)\bar{\psi}^i\psi^i(r',\tau)\nonumber
\end{eqnarray}

\begin{figure}[t]
\begin{center}
\begin{tabular}{cc}
\includegraphics[width=0.1\textwidth]{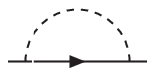} & \includegraphics[width=0.1\textwidth]{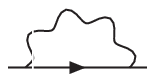}
\end{tabular}
\end{center}
\caption{The self energy diagrams to leading non-trivial order in
$\{\Delta_{\phi}$, $\Delta_A\}$ (left) and $\alpha$ (right).}
\label{fig:selfenergyDiag}
\end{figure}

As usual, we assume large momentum cutoff $\Lambda$ for the above
fermion modes, and perform the renormalization of the bare coupling
constants $e^2$,$v_F$,$\Delta_{\phi}$,$\Delta_A$\cite{amitBook}. To
first order we find that the imaginary time Greens function
(Fig.(\ref{fig:selfenergyDiag})) satisfies
\begin{eqnarray}
 G^{-1}_{i\omega}(\bf
k)&=&-i\omega\left(1+\frac{\Delta_{\phi}}{2\pi}\log\frac{\Lambda}{|\omega|}+\frac{\Delta_A}{\pi}\log\frac{\Lambda}{|\omega|}\right)
\nonumber\\
&+&\left(v_F+\frac{e^2}{4}\log\frac{\Lambda}{k}\right)\sigma\cdot{\bf
k}
\end{eqnarray}
The renormalization condition demands that we absorb the dependence
on the cut-off $\Lambda$ into a field rescaling constant $Z$ and the
bare couplings. We do so at an arbitrary scale $\omega=k=\kappa$
where we demand that
\begin{eqnarray}  G_{i\omega}(\bk)|_{\kappa}= Z
G^R_{i\omega}(\bk)|_{\kappa}=Z(-i\kappa+v^R_F\sigma\cdot\kappa)^{-1}
\end{eqnarray}
The renormalized Greens function $G^R_{i\omega}(\bk)$, {\it at any}
$\omega$ {\it and} $\bk$, is now independent of $\Lambda$. This
leads to the RG equation for the Fermi velocity
\begin{eqnarray}\label{eq:BetafxnsVF}  \beta_{v_F}=\frac{\partial
v_F}{\partial\log\Lambda}=
v_F\left(\frac{\Delta_{\phi}}{2\pi}+\frac{\Delta_{A}}{\pi}-\frac{e^2}{4v_F}\right)
\end{eqnarray}
\begin{figure}[h]
\begin{center}
\begin{tabular}{cccccc}
\includegraphics[width=0.06\textwidth]{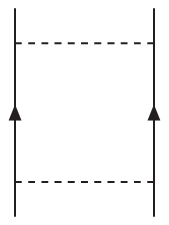} &
\includegraphics[width=0.06\textwidth]{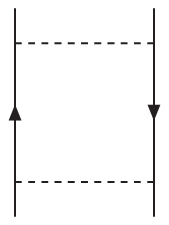}&
\includegraphics[width=0.08\textwidth]{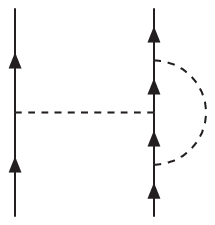} &
\includegraphics[width=0.08\textwidth]{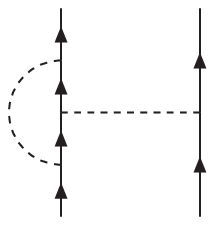} &
\includegraphics[width=0.08\textwidth]{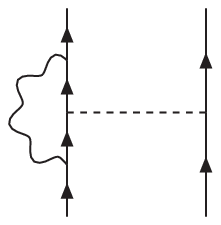} &
\includegraphics[width=0.08\textwidth]{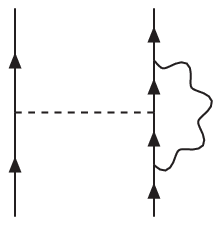}
\end{tabular}
\end{center}
\caption{Diagrams which contribute to the disorder renormalization.}
\label{fig:disorderRG}
\end{figure}

To determine the RG scaling of the disorder variances we need to
analyze the $\beta$ functions of the effective replica coupling
constants. We do so by writing the equations for the irreducible
four point vertex: $ \Gamma^{(4)}=\frac{G_4}{G^4_2}=\langle
\psi^i\bar{\psi}^i\psi^j\bar{\psi}^j\rangle_{con,amp}$. The
renormalization prescription demands that at a scale
$\omega=k=\kappa$, $\Gamma^{(4)}=\frac{1}{Z^2}\Gamma_R^{(4)} $. This
means that at arbitrary $\omega,{\bf k}$, the quantity
$Z^2\Gamma^{(4)}$ can be made independent of $\Lambda$. To this
order in coupling constants (Fig. (\ref{fig:disorderRG})) we find
\begin{eqnarray} \beta_{\Delta_{\phi}} &=& \frac{\partial
\Delta_{\phi}}{\partial
\log\Lambda}=-2\Delta_{\phi}\left(\frac{\Delta_{\phi}}{2\pi}+\frac{\Delta_A}{\pi}-\frac{e^2}{4v_F}\right)\label{eq:BetafxnsDeltaPhi}\\
\beta_{\Delta_A}&=&\frac{\partial \Delta_{A}}{\partial \log\Lambda}=0\label{eq:BetafxnsDeltaA}\\
\beta_{e^2}&=&\frac{\partial e^2}{\partial
\log\Lambda}=0\label{eq:BetafxnsDelta_eSq},\end{eqnarray} which
agrees with Ref.\cite{fosterAleinerPRB2008}. As argued in
Ref.\cite{HerbutJuricicVafekPRL2008} the last equation is exact. The
corresponding flow diagram is shown in Fig.\ref{fig:RGflows}.

Defining the dimensionless Coulomb coupling constant
$\alpha=e^2/\eps v_F$, the above equations imply the existence of
two RG invariants $\gamma=\Delta_{\phi}/\alpha^2$ and $\Delta_A$,
i.e.
\begin{equation}  \frac{\partial \gamma}{\partial
\log\Lambda}=\frac{\partial \Delta_{A}}{\partial
\log\Lambda}=0.\end{equation}

Since conductivity does not acquire anomalous dimension we have
\begin{eqnarray}  \left(\frac{\partial}{\partial
\log\Lambda}+\hat{\mathcal{B}}\right)
\sigma(\omega;\Lambda,\Delta_{\phi},\Delta_{A},v_F,e^2)=0,
\end{eqnarray}
where the differential operator
\begin{eqnarray}
\hat{\mathcal{B}}=\beta_{\Delta_{\phi}}\frac{\partial}{\partial
\Delta_{\phi}}+\beta_{\Delta_{A}}\frac{\partial}{\partial
\Delta_{A}}+\beta_{v_F}\frac{\partial}{\partial
v_F}+\beta_{e^2}\frac{\partial}{\partial e^2}.
\end{eqnarray}
The solution of the above RG equation must satisfy the scaling law
\begin{eqnarray}\label{eq:sigmaCalSym}
&&\sigma(\omega;\rho\Lambda,\Delta_{\phi}(\rho\Lambda),
\Delta_{A}(\rho\Lambda),v_F(\rho\Lambda),e^2(\rho\Lambda))=\nonumber\\
&&\sigma(\omega;\Lambda,\Delta_{\phi}(\Lambda),\Delta_{A}(\Lambda),v_F(\Lambda),e^2(\Lambda))
\end{eqnarray}
where $\rho$ is a positive real number \cite{amitBook}.

%
\begin{figure}[t]
\begin{center}
\begin{tabular}{ccccc}
\includegraphics[width=0.09\textwidth]{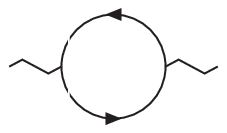} &
\includegraphics[width=0.09\textwidth]{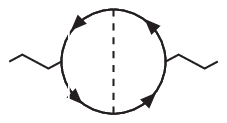}&
\includegraphics[width=0.09\textwidth]{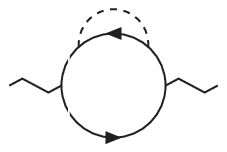} &
\includegraphics[width=0.09\textwidth]{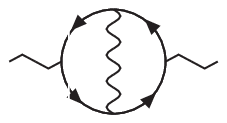} &
\includegraphics[width=0.09\textwidth]{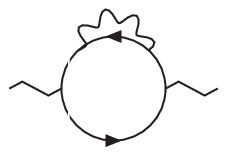}
\end{tabular}
\end{center}
\caption{Diagrammatic contribution to the (a.c.) electrical
conductivity within Kubo formula. From left to right: free Dirac
fermion contribution, disorder vertex and self energy corrections,
Coulomb interaction vertex and self energy contribution.}
\label{fig:conductivity}
\end{figure}
\begin{figure}[t]
\begin{center}
\begin{tabular}{c}
\includegraphics[width=0.5\textwidth]{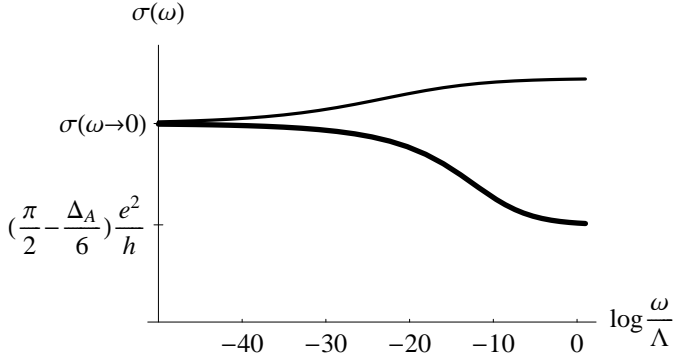}
\end{tabular}
\end{center}
\caption{Illustrative dependence of $\sigma$ on frequency for
$\Delta_A=1$ and $\gamma=0.95\frac{\pi^2}{32\Delta}$. The lower
curve corresponds to the starting condition $\alpha\approx 0$ (near
the repulsive fixed point at the origin of Fig.(\ref{fig:RGflows}))
and the upper to $\alpha\approx 3.25$ (near the strong coupling
repulsive fixed point). Notice that since both curves asymptote to
the same fixed point, the $\omega\rightarrow 0$ limit of $\sigma$ is
the same and the value is given by Eq.(\ref{eq:Conductivity}).}
\label{fig:sigmaflows}
\end{figure}

A pedestrian perturbation theory calculation to the leading order in
coupling constants (Fig.\ref{fig:conductivity}) gives
\begin{eqnarray}\label{eq:sigmaPert}
\sigma_{pt}(\omega)=4\frac{e^2}{h}\left[\frac{\pi}{8}-\frac{\Delta_{\phi}}{24}-
\frac{\Delta_{A}}{24}+\alpha\frac{\pi}{16}\left(\frac{25}{6}-\pi\right)\right].
\end{eqnarray}
This extends the result found in \cite{HerbutJuricicVafekPRL2008} to
include the scalar-disorder potential contribution. Note that the
above expression does not satisfy the scaling law
(\ref{eq:sigmaCalSym}), since according to
Eqs.(\ref{eq:BetafxnsVF}-\ref{eq:BetafxnsDelta_eSq}) the coupling
constant $\alpha$ and the scalar disorder variance $\Delta_{\phi}$
do have non-trivial dependence on $\Lambda$. Nevertheless, to the
same order in the coupling constants, both (\ref{eq:sigmaCalSym})
and (\ref{eq:sigmaPert}) can be satisfied if
\begin{eqnarray}
\sigma(\omega)=4\frac{e^2}{h}\left[\frac{\pi}{8}-\frac{\gamma\alpha^2(\frac{\omega}{\Lambda})}{24}
-\frac{\Delta_{A}}{24}+\frac{\pi\alpha(\frac{\omega}{\Lambda})}{16}\left(\frac{25}{6}-\pi\right)\right].
\end{eqnarray}
where we used $\Delta_{\phi}(\rho)=\gamma\alpha^2(\rho)$. The
dimensionless Coulomb coupling constant $\alpha(\rho)$ is defined as
the solution of
\begin{eqnarray}\label{eq:BetafxnsDelta_alpha}
\frac{\partial \alpha}{\partial \log
\rho}=-\alpha\left(\frac{\gamma}{2\pi}\alpha^2-\frac{\alpha}{4}+\frac{\Delta_A}{\pi}\right)
\end{eqnarray}
with the initial condition $\alpha(1)=\alpha$. The functional
dependence can be found implicitly
\begin{eqnarray}&&
\frac{\frac{4\gamma}{\pi}\alpha(\frac{\omega}{\Lambda})-(1+A)}{\frac{4\gamma}{\pi}\alpha(\frac{\omega}{\Lambda})-(1-A)}
\left(\frac{\alpha^2(\frac{\omega}{\Lambda})}{(\frac{4\gamma}{\pi}\alpha(\frac{\omega}{\Lambda})-1)^2-A^2}\right)^A=\nonumber\\
&&\frac{\frac{4\gamma}{\pi}\alpha-(1+A)}{\frac{4\gamma}{\pi}\alpha-(1-A)}
\left(\frac{\alpha^2}{(\frac{4\gamma}{\pi}\alpha-1)^2-A^2}\right)^A\left(\frac{\omega}{\Lambda}\right)^{\frac{\pi}{16\gamma}A(A^2-1)},
\end{eqnarray}
where $ A=\sqrt{1-\frac{32}{\pi^2}\gamma\Delta_A}.$ The above
equation is easily inverted numerically. As an illustration, the
resulting conductivity as a function of frequency for $\Delta_A=1$
and in the vicinity of the multicritical trajectory is plotted in
Fig (\ref{fig:sigmaflows}).

The explicit dependence on $\omega$ can be found in some limiting
cases. In the vicinity of the IR stable fixed line, but away from
the multicritical point (see Fig.\ref{fig:RGflows}), we find
\begin{eqnarray}
\alpha\left(\frac{\omega}{\Lambda}\right)\!\!&\approx&\!\!\frac{\pi}{4\gamma}\frac{1-\sqrt{1-\frac{32}{\pi^2}\gamma\Delta_A}}{1-\left[1-\frac{\pi}{4\gamma\alpha}
\left(1-\sqrt{1-\frac{32}{\pi^2}\gamma\Delta_A}\right)\right]\left(\frac{\omega}{\Lambda}\right)^{\theta}}\nonumber\\
\end{eqnarray}
where the (crossover) exponent
\begin{eqnarray}
\theta=\frac{\pi}{16\gamma}\sqrt{1-\frac{32}{\pi^2}\gamma\Delta_A}\left(1-\sqrt{1-\frac{32}{\pi^2}\gamma\Delta_A}\right)\end{eqnarray}

Note also that
\begin{eqnarray}
\frac{\partial \sigma(\omega)}{\partial
\log\omega}&=&4\frac{e^2}{h}\left[\frac{\pi}{16}\left(\frac{25}{6}-\pi\right)-\frac{\gamma}{12}\alpha(\frac{\omega}{\Lambda})\right]\frac{\partial
\alpha}{\partial \log\omega}\nonumber.
\end{eqnarray}
The Eq.(\ref{eq:BetafxnsDelta_alpha}) then implies that for initial
$\alpha<\frac{3\pi}{4\gamma}\left(\frac{25}{6}-\pi\right)$, the
conductivity in the Region I to the left of the IR fixed line (light
grey shaded portion of Region I in Fig.\ref{fig:RGflows}) increases
with decreasing $\omega$. On the other hand, to the right of the IR
fixed line (dark shaded portion of Region I) $\sigma(\omega)$
decreases with decreasing $\omega$. In either case, however, as
$\omega\rightarrow 0$, $\sigma$ asymptotes to the value given by
Eq.(\ref{eq:Conductivity}). Moreover, while in the latter case, the
large frequency limit is outside of the scope of perturbative RG, in
the former case the high frequency limit of conductivity is
$\frac{e^2}{h}\left(\frac{\pi}{2}-\frac{\Delta_A}{6}\right).$

Thus, the appearance of the infra-red (locally) stable line of fixed
points at finite Coulomb coupling and finite disorder (Fig.
\ref{fig:RGflows}) provides a natural theoretical avenue towards
non-universality of the longitudinal electrical conductivity, since
the precise position along such line is typically beyond
experimental control. Nevertheless, any correlation between the
variance of the (independently measurable) strain configurations and
the minimal conductivity would provide a good test of the above
theory. It is also important to address the effects of the general
form of disorder. As argued in Ref.\cite{fosterAleinerPRB2008}, if
one starts with the most general disorder potential allowed by the
symmetry of the graphene honeycomb lattice, the RG flow trajectories
runaway to strong coupling/strong disorder and the physics is
perturbatively untractable. The picture presented here may
nevertheless be physically relevant for suspended samples, since the
primary mode of coupling of the smooth deformations to the graphene
Dirac fermions is via scalar and vector potentials, and all but the
absence of the short wavelength components of the scattering
potential gives the above (globally unstable) fixed line a large
basin of influence.

I wish to acknowledge the hospitality of the Aspen Center for
Physics where part of this work was completed.

\bibliography{rs}

\begin{thebibliography}{24}
\expandafter\ifx\csname natexlab\endcsname\relax\def\natexlab#1{#1}\fi
\expandafter\ifx\csname bibnamefont\endcsname\relax
  \def\bibnamefont#1{#1}\fi
\expandafter\ifx\csname bibfnamefont\endcsname\relax
  \def\bibfnamefont#1{#1}\fi
\expandafter\ifx\csname citenamefont\endcsname\relax
  \def\citenamefont#1{#1}\fi
\expandafter\ifx\csname url\endcsname\relax
  \def\url#1{\texttt{#1}}\fi
\expandafter\ifx\csname urlprefix\endcsname\relax\def\urlprefix{URL }\fi
\providecommand{\bibinfo}[2]{#2}
\providecommand{\eprint}[2][]{\url{#2}}

\bibitem[{\citenamefont{Novoselov et~al.}(2005)\citenamefont{Novoselov, Geim,
  Morozov, Jiang, Katsnelson, Grigorieva, Dubonos, and
  Firsov}}]{Novoselov:2005fk}
\bibinfo{author}{\bibfnamefont{K.~S.} \bibnamefont{Novoselov}},
  \bibinfo{author}{\bibfnamefont{A.~K.} \bibnamefont{Geim}},
  \bibinfo{author}{\bibfnamefont{S.~V.} \bibnamefont{Morozov}},
  \bibinfo{author}{\bibfnamefont{D.}~\bibnamefont{Jiang}},
  \bibinfo{author}{\bibfnamefont{M.~I.} \bibnamefont{Katsnelson}},
  \bibinfo{author}{\bibfnamefont{I.~V.} \bibnamefont{Grigorieva}},
  \bibinfo{author}{\bibfnamefont{S.~V.} \bibnamefont{Dubonos}},
  \bibnamefont{and} \bibinfo{author}{\bibfnamefont{A.~A.}
  \bibnamefont{Firsov}}, \bibinfo{journal}{Nature}
  \textbf{\bibinfo{volume}{438}}, \bibinfo{pages}{197} (\bibinfo{year}{2005}).

\bibitem[{\citenamefont{Zhang et~al.}(2005)\citenamefont{Zhang, Tan, Stormer,
  and Kim}}]{Zhang:2005uq}
\bibinfo{author}{\bibfnamefont{Y.}~\bibnamefont{Zhang}},
  \bibinfo{author}{\bibfnamefont{Y.-W.} \bibnamefont{Tan}},
  \bibinfo{author}{\bibfnamefont{H.~L.} \bibnamefont{Stormer}},
  \bibnamefont{and} \bibinfo{author}{\bibfnamefont{P.}~\bibnamefont{Kim}},
  \bibinfo{journal}{Nature} \textbf{\bibinfo{volume}{438}},
  \bibinfo{pages}{201} (\bibinfo{year}{2005}).

\bibitem[{\citenamefont{Geim and Novoselov}(2007)}]{Geim:2007lr}
\bibinfo{author}{\bibfnamefont{A.~K.} \bibnamefont{Geim}} \bibnamefont{and}
  \bibinfo{author}{\bibfnamefont{K.~S.} \bibnamefont{Novoselov}},
  \bibinfo{journal}{Nat. Mater.} \textbf{\bibinfo{volume}{6}},
  \bibinfo{pages}{183} (\bibinfo{year}{2007}).

\bibitem[{\citenamefont{Herbut et~al.}(2008)\citenamefont{Herbut,
  Juri\v{c}i\'{c}, and Vafek}}]{HerbutJuricicVafekPRL2008}
\bibinfo{author}{\bibfnamefont{I.~F.} \bibnamefont{Herbut}},
  \bibinfo{author}{\bibfnamefont{V.}~\bibnamefont{Juri\v{c}i\'{c}}},
  \bibnamefont{and} \bibinfo{author}{\bibfnamefont{O.}~\bibnamefont{Vafek}},
  \bibinfo{journal}{\prl} \textbf{\bibinfo{volume}{100}}, \bibinfo{eid}{046403}
  (\bibinfo{year}{2008}).

\bibitem[{\citenamefont{Gonzales et~al.}(1994)\citenamefont{Gonzales, Guinea,
  and Vozmediano}}]{Gonzales1994}
\bibinfo{author}{\bibfnamefont{J.}~\bibnamefont{Gonzales}},
  \bibinfo{author}{\bibfnamefont{F.}~\bibnamefont{Guinea}}, \bibnamefont{and}
  \bibinfo{author}{\bibfnamefont{M.~A.~H.} \bibnamefont{Vozmediano}},
  \bibinfo{journal}{Nucl. Phys. B} \textbf{\bibinfo{volume}{424}},
  \bibinfo{pages}{595} (\bibinfo{year}{1994}).

\bibitem[{\citenamefont{Herbut}(2006)}]{herbutPRL06}
\bibinfo{author}{\bibfnamefont{I.~F.} \bibnamefont{Herbut}},
  \bibinfo{journal}{\prl} \textbf{\bibinfo{volume}{97}}, \bibinfo{eid}{146401}
  (\bibinfo{year}{2006}).

\bibitem[{\citenamefont{Vafek}(2007)}]{vafek:prl2007}
\bibinfo{author}{\bibfnamefont{O.}~\bibnamefont{Vafek}},
  \bibinfo{journal}{\prl} \textbf{\bibinfo{volume}{98}}, \bibinfo{eid}{216401}
  (\bibinfo{year}{2007}).

\bibitem[{\citenamefont{Sheehy and Schmalian}(2007)}]{sheehy:226803}
\bibinfo{author}{\bibfnamefont{D.~E.} \bibnamefont{Sheehy}} \bibnamefont{and}
  \bibinfo{author}{\bibfnamefont{J.}~\bibnamefont{Schmalian}},
  \bibinfo{journal}{\prl} \textbf{\bibinfo{volume}{99}}, \bibinfo{eid}{226803}
  (\bibinfo{year}{2007}).

\bibitem[{\citenamefont{Fritz et~al.}(2008)\citenamefont{Fritz, Schmalian,
  M\"{u}ller, and Sachdev}}]{fritz:085416}
\bibinfo{author}{\bibfnamefont{L.}~\bibnamefont{Fritz}},
  \bibinfo{author}{\bibfnamefont{J.}~\bibnamefont{Schmalian}},
  \bibinfo{author}{\bibfnamefont{M.}~\bibnamefont{M\"{u}ller}},
  \bibnamefont{and} \bibinfo{author}{\bibfnamefont{S.}~\bibnamefont{Sachdev}},
  \bibinfo{journal}{\prb} \textbf{\bibinfo{volume}{78}}, \bibinfo{eid}{085416}
  (\bibinfo{year}{2008}).

\bibitem[{\citenamefont{Fisher et~al.}(1990)\citenamefont{Fisher, Grinstein,
  and Girvin}}]{Fisher1990}
\bibinfo{author}{\bibfnamefont{M.~P.~A.} \bibnamefont{Fisher}},
  \bibinfo{author}{\bibfnamefont{G.}~\bibnamefont{Grinstein}},
  \bibnamefont{and} \bibinfo{author}{\bibfnamefont{S.~M.}
  \bibnamefont{Girvin}}, \bibinfo{journal}{Phys. Rev. Lett.}
  \textbf{\bibinfo{volume}{64}}, \bibinfo{pages}{587} (\bibinfo{year}{1990}).

\bibitem[{\citenamefont{Cha et~al.}(1991)\citenamefont{Cha, Fisher, Girvin,
  Wallin, and Young}}]{Cha1991}
\bibinfo{author}{\bibfnamefont{M.-C.} \bibnamefont{Cha}},
  \bibinfo{author}{\bibfnamefont{M.~P.~A.} \bibnamefont{Fisher}},
  \bibinfo{author}{\bibfnamefont{S.~M.} \bibnamefont{Girvin}},
  \bibinfo{author}{\bibfnamefont{M.}~\bibnamefont{Wallin}}, \bibnamefont{and}
  \bibinfo{author}{\bibfnamefont{A.~P.} \bibnamefont{Young}},
  \bibinfo{journal}{Phys. Rev. B} \textbf{\bibinfo{volume}{44}},
  \bibinfo{pages}{6883} (\bibinfo{year}{1991}).

\bibitem[{\citenamefont{Damle and Sachdev}(1997)}]{Damle97}
\bibinfo{author}{\bibfnamefont{K.}~\bibnamefont{Damle}} \bibnamefont{and}
  \bibinfo{author}{\bibfnamefont{S.}~\bibnamefont{Sachdev}},
  \bibinfo{journal}{\prb} \textbf{\bibinfo{volume}{56}}, \bibinfo{pages}{8714}
  (\bibinfo{year}{1997}).

\bibitem[{\citenamefont{Nair et~al.}(2008)\citenamefont{Nair, Blake,
  Grigorenko, Novoselov, Booth, Stauber, Peres, and Geim}}]{Nair06062008}
\bibinfo{author}{\bibfnamefont{R.~R.} \bibnamefont{Nair}},
  \bibinfo{author}{\bibfnamefont{P.}~\bibnamefont{Blake}},
  \bibinfo{author}{\bibfnamefont{A.~N.} \bibnamefont{Grigorenko}},
  \bibinfo{author}{\bibfnamefont{K.~S.} \bibnamefont{Novoselov}},
  \bibinfo{author}{\bibfnamefont{T.~J.} \bibnamefont{Booth}},
  \bibinfo{author}{\bibfnamefont{T.}~\bibnamefont{Stauber}},
  \bibinfo{author}{\bibfnamefont{N.~M.~R.} \bibnamefont{Peres}},
  \bibnamefont{and} \bibinfo{author}{\bibfnamefont{A.~K.} \bibnamefont{Geim}},
  \bibinfo{journal}{Science} \textbf{\bibinfo{volume}{320}},
  \bibinfo{pages}{1308} (\bibinfo{year}{2008}).

\bibitem[{\citenamefont{Li et~al.}(2008)\citenamefont{Li, Henriksen, Jiang,
  Hao, Martin, Kim, Stormer, and Basov}}]{basov2008}
\bibinfo{author}{\bibfnamefont{Z.~Q.} \bibnamefont{Li}},
  \bibinfo{author}{\bibfnamefont{E.~A.} \bibnamefont{Henriksen}},
  \bibinfo{author}{\bibfnamefont{Z.}~\bibnamefont{Jiang}},
  \bibinfo{author}{\bibfnamefont{Z.}~\bibnamefont{Hao}},
  \bibinfo{author}{\bibfnamefont{M.~C.} \bibnamefont{Martin}},
  \bibinfo{author}{\bibfnamefont{P.}~\bibnamefont{Kim}},
  \bibinfo{author}{\bibfnamefont{H.~L.} \bibnamefont{Stormer}},
  \bibnamefont{and} \bibinfo{author}{\bibfnamefont{D.~N.} \bibnamefont{Basov}},
  \bibinfo{journal}{Nature Physics} \textbf{\bibinfo{volume}{4}},
  \bibinfo{pages}{532} (\bibinfo{year}{2008}).

\bibitem[{\citenamefont{Dawlaty et~al.}(2008)\citenamefont{Dawlaty, Shivaraman,
  Strait, George, Chandrashekhar, Rana, Spencer, Veksler, and
  Chen}}]{dawlaty-2008}
\bibinfo{author}{\bibfnamefont{J.~M.} \bibnamefont{Dawlaty}},
  \bibinfo{author}{\bibfnamefont{S.}~\bibnamefont{Shivaraman}},
  \bibinfo{author}{\bibfnamefont{J.}~\bibnamefont{Strait}},
  \bibinfo{author}{\bibfnamefont{P.}~\bibnamefont{George}},
  \bibinfo{author}{\bibfnamefont{M.}~\bibnamefont{Chandrashekhar}},
  \bibinfo{author}{\bibfnamefont{F.}~\bibnamefont{Rana}},
  \bibinfo{author}{\bibfnamefont{M.~G.} \bibnamefont{Spencer}},
  \bibinfo{author}{\bibfnamefont{D.}~\bibnamefont{Veksler}}, \bibnamefont{and}
  \bibinfo{author}{\bibfnamefont{Y.}~\bibnamefont{Chen}}
  (\bibinfo{year}{2008}), \bibinfo{note}{arXiv.org:0801.3302}.

\bibitem[{\citenamefont{Suzuura and Ando}(2002)}]{suzuura}
\bibinfo{author}{\bibfnamefont{H.}~\bibnamefont{Suzuura}} \bibnamefont{and}
  \bibinfo{author}{\bibfnamefont{T.}~\bibnamefont{Ando}},
  \bibinfo{journal}{\prb} \textbf{\bibinfo{volume}{65}},
  \bibinfo{pages}{235412} (\bibinfo{year}{2002}).

\bibitem[{\citenamefont{Manes}(2007)}]{manes:045430}
\bibinfo{author}{\bibfnamefont{J.~L.} \bibnamefont{Manes}},
  \bibinfo{journal}{\prb} \textbf{\bibinfo{volume}{76}}, \bibinfo{eid}{045430}
  (\bibinfo{year}{2007}).

\bibitem[{\citenamefont{Mariani and von Oppen}(2008)}]{mariani:076801}
\bibinfo{author}{\bibfnamefont{E.}~\bibnamefont{Mariani}} \bibnamefont{and}
  \bibinfo{author}{\bibfnamefont{F.}~\bibnamefont{von Oppen}},
  \bibinfo{journal}{\prl} \textbf{\bibinfo{volume}{100}}
  (\bibinfo{year}{2008}).

\bibitem[{\citenamefont{Ludwig et~al.}(1994)\citenamefont{Ludwig, Fisher,
  Shankar, and Grinstein}}]{Ludwig:prb94}
\bibinfo{author}{\bibfnamefont{A.}~\bibnamefont{Ludwig}},
  \bibinfo{author}{\bibfnamefont{M.~P.~A.} \bibnamefont{Fisher}},
  \bibinfo{author}{\bibfnamefont{R.}~\bibnamefont{Shankar}}, \bibnamefont{and}
  \bibinfo{author}{\bibfnamefont{G.}~\bibnamefont{Grinstein}},
  \bibinfo{journal}{\prb} \textbf{\bibinfo{volume}{50}}, \bibinfo{pages}{7526}
  (\bibinfo{year}{1994}).

\bibitem[{\citenamefont{Aleiner and Efetov}(2006)}]{aleinerEfetov2006}
\bibinfo{author}{\bibfnamefont{I.~L.} \bibnamefont{Aleiner}} \bibnamefont{and}
  \bibinfo{author}{\bibfnamefont{K.~B.} \bibnamefont{Efetov}},
  \bibinfo{journal}{\prl} \textbf{\bibinfo{volume}{97}}, \bibinfo{eid}{236801}
  (\bibinfo{year}{2006}).

\bibitem[{\citenamefont{Vafek and Case}(2008)}]{VafekCasePRB2008}
\bibinfo{author}{\bibfnamefont{O.}~\bibnamefont{Vafek}} \bibnamefont{and}
  \bibinfo{author}{\bibfnamefont{M.~J.} \bibnamefont{Case}},
  \bibinfo{journal}{\prb} \textbf{\bibinfo{volume}{77}}, \bibinfo{eid}{033410}
  (\bibinfo{year}{2008}).

\bibitem[{\citenamefont{Foster and Aleiner}(2008)}]{fosterAleinerPRB2008}
\bibinfo{author}{\bibfnamefont{M.~S.} \bibnamefont{Foster}} \bibnamefont{and}
  \bibinfo{author}{\bibfnamefont{I.~L.} \bibnamefont{Aleiner}},
  \bibinfo{journal}{\prb} \textbf{\bibinfo{volume}{77}}, \bibinfo{eid}{195413}
  (\bibinfo{year}{2008}).

\bibitem[{\citenamefont{Guinea et~al.}(2008)\citenamefont{Guinea, Horovitz, and
  Doussal}}]{guinea:205421}
\bibinfo{author}{\bibfnamefont{F.}~\bibnamefont{Guinea}},
  \bibinfo{author}{\bibfnamefont{B.}~\bibnamefont{Horovitz}}, \bibnamefont{and}
  \bibinfo{author}{\bibfnamefont{P.~L.} \bibnamefont{Doussal}},
  \bibinfo{journal}{\prb} \textbf{\bibinfo{volume}{77}} (\bibinfo{year}{2008}).

\bibitem[{\citenamefont{Amit and Martin-Mayor}(2005)}]{amitBook}
\bibinfo{author}{\bibfnamefont{D.~J.} \bibnamefont{Amit}} \bibnamefont{and}
  \bibinfo{author}{\bibfnamefont{V.}~\bibnamefont{Martin-Mayor}},
  \emph{\bibinfo{title}{Field Theory, the Renormalization Group and Critical
  Phenomena}} (\bibinfo{publisher}{World Scientific},
  \bibinfo{address}{Singapore}, \bibinfo{year}{2005}), \bibinfo{edition}{3rd}
  ed., \bibinfo{note}{p.214}.

\end{thebibliography}
\end{document}